# A True Real-Time Success Story:
# the Case of Collecting Beauty-ful Data
# at the LHCb Experiment

Federico Alessio, on behalf of the LHCb Collaboration

*Abstract*—**The LHCb experiment at CERN is currently completing its first big data taking campaign at the LHC started in 2009. It has been collecting data at more than 2.5 times its nominal design luminosity value and with a global efficiency of ~92%. Even more striking, the efficiency between online and offline recorded luminosity, obtained by comparing the data quality output, is close to 99%, which highlights how well the detector, its data acquisition system and its control system have been performing despite much harsher and more variable conditions than initially foreseen.**

**In this paper, the excellent performance of the LHCb experiment will be described, by transversally tying together the timing and data acquisition system, the software trigger, the real-time calibration and the shifters interaction with the control system. Particular attention will be given to their real-time aspects, which allow performing an online reconstruction that is at the same performance level as the offline one through a real-time calibration and alignment of the full detector. In addition, the various solutions that have been chosen to operate the experiment safely and synchronously with the various phases of the LHC operations will also be shown. In fact, the quasi-autonomous control system of the LHCb experiment is the key to explain how such a large detector can be operated successfully around the clock by a pool of ~300 non-expert shifters. Finally, a critical review of the experiment will be presented in order to justify the reasons for a major upgrade of the detector.**

## I. INTRODUCTION

THE LHCb experiment at CERN (Fig. 1) is dedicated to precision measurements of heavy flavour physics with the main aim to probe physics beyond the Standard Model, by studying very rare decays of beauty and charm-flavoured hadrons, and by measuring precisely *CP*-violating observables [1].

In order to successfully satisfy its physics reach in a challenging environment such as the CERN Large Hadron Collider, it is paramount that the LHCb detector performs with the highest possible degree of precision, reliability and efficiency, collecting as much data as possibly delivered by the accelerator while still being operated by non-expert shifters around the clock. In almost a decade of operations, from 2009 onwards, the LHCb experiment collected a total of ~7 fb$^{-1}$ with a global efficiency of ~92% at the date of the publication of the proceedings (Fig. 2), with an additional 2 fb$^{-1}$ foreseen to be recorded in the last year of operation.

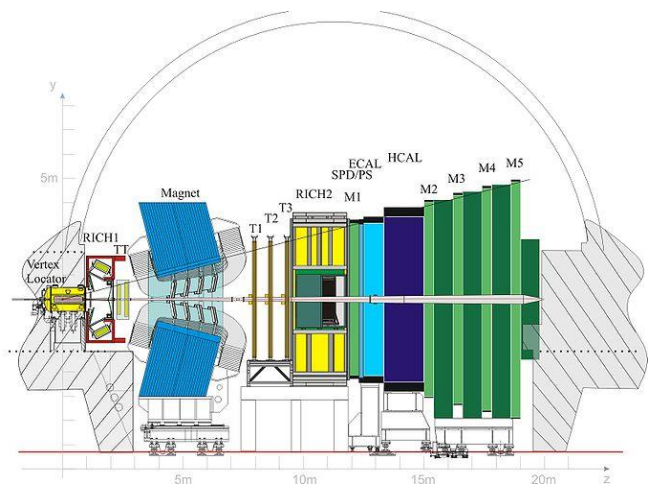

Fig. 1: The LHCb detector at CERN. The detector is built to perform precise vertexing, tracking, particle identification, calorimetry and muon detection. It is also equipped with a first-level hardware trigger that bases its decision on multiplicity, energy and muon detection.

While the operational efficiency – defined as the pure ratio between the collected and delivered luminosity from the LHC accelerator – is mostly driven by the readout system capabilities (e.g. deadtime), the actual online/offline efficiency, obtained by comparing the amount of data collected and the amount of data actually saved to tape without being further rejected, is ~99%. This shows that the detector is behaving at the same performance as the offline reconstruction. The key to the success behind such excellent performance are to be found in various closely intertwined aspects of the systems and such aspects are the subject of this paper. These are:

- automatized operations
- centralized readout supervision
- a powerful, flexible and efficient trigger system
- the real-time calibration and alignment procedures

F. Alessio is with CERN, Rue de Meyrin, 1211 Geneva, Switzerland (e-mail: federico.alessio@cern.ch).

- online monitoring and control of luminosity, beam and background conditions.

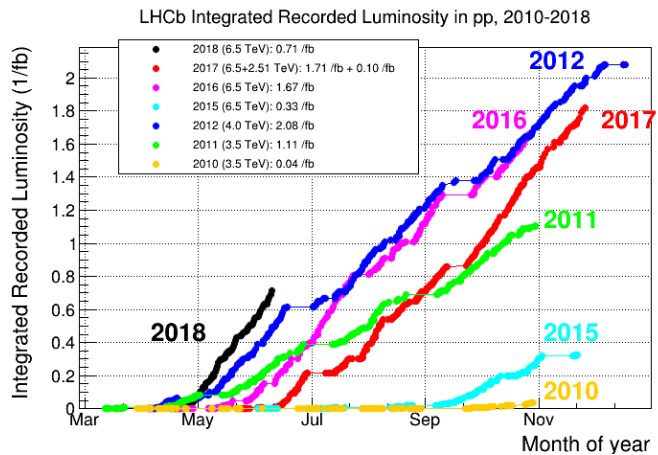

Fig. 2: Amount of data recorded by the LHCb experiment in the first decade of its operation.

Such kind of implementations allowed the experiment to be operated routinely on a daily-basis by a pool of ~300 non-expert shifters, with minimal training and well below 1% operational efficiency loss due to human errors. The average number of shifts per trained shifter is in fact only ~6/year.

In this paper, each of the above mentioned aspects are briefly reviewed in order to highlight the aspects of data taking that made the LHCb experiment a true success of its real-time approach.

## II. AUTOMATIZED OPERATIONS

The LHCb detector can be operated in a quasi-automatic way: most of the procedures and the activities are supervised by an Experiment Control System (ECS) implemented via a SCADA system [2]. The global LHCb ECS system is able to perform or suggest to the shifter the actions that need to be done to perform a routine in a specific condition, for example start data acquisition when in "Physics data taking mode" or switch HV OFF when in an unsafe mode. The shifter is only asked to supervise that the software performs as required and the shifter is required to acknowledge the operation - for those cases in which acknowledge is required – to make sure that he/she is informed regarding the action being taken.

Among other reasons, such kind of automation is possible thanks to the fact that the LHCb ECS is tightly intertwined to the hardware and software state of the LHC accelerator. An extensive data exchange framework (Fig. 3) with the LHC has been developed such that it is possible to receive and archive the most important data regarding the accelerator and at the same time transmit feedback to the LHC regarding operations in LHCb [3]. This allows to perform real-time actions based on a predefined matrix of conditions.

As a practical example, it is possible to inform the LHCb ECS about the state of the accelerator (i.e. "*INJECTION*") and at the same time the LHCb ECS suggests to the shifter what correct voltage state for each sub-detector for that specific accelerator state. If by any reason, something is out of the foreseen correct final state, an alarm and/or a sound signals the problem to the shifter. In addition, if the problem impacts safe operations of the experiment the information regarding a problem is also transmitted to the LHC and in the extreme case inhibit further operations (practically forbid injection).

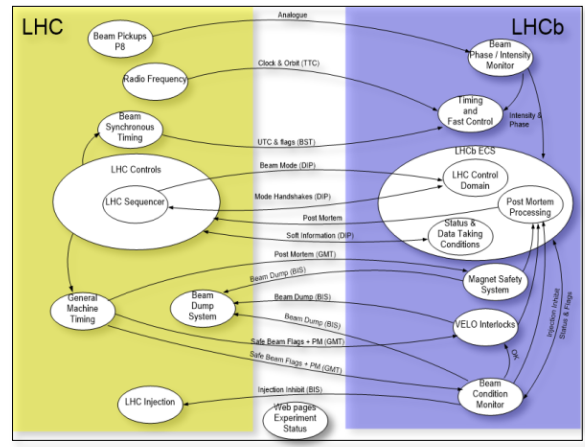

Fig. 3: Schematic representation of the data exchange between the LHC accelerator and the LHCb detector.

Inhibits or vetoes are ultimately done through hardware interlocks directly connected to the LHC Beam Interlock System (BIS) for higher degrees of safety. Such type of communication is a standard way of operating the experiment and it is part of daily routine actions.

A screenshot of the control system panel that the shifter is required to use in the LHCb control room to check the state of the LHCb detector and all of its sub-systems with respect to the state of the LHC accelerator is shown in Fig. 4. Such kind of panels is the panel used to act on the global system by the central shifter crew in the LHCb control room.

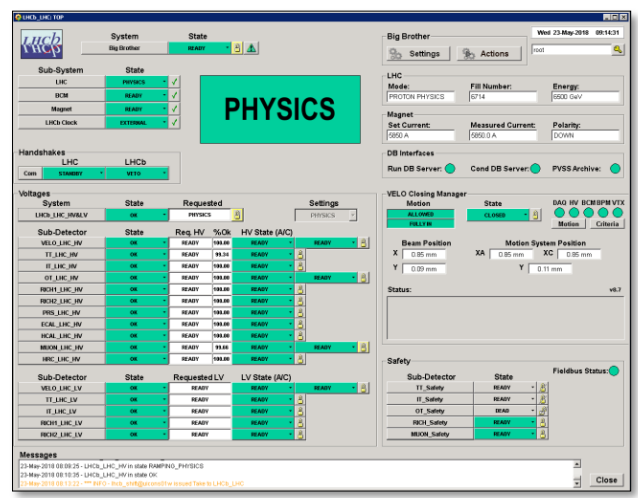

Fig. 4: The control room panel that is made available to the shifter for monitoring and operating the LHCb detector. It summarizes the state of the LHCb detector, of the LHC accelerator, the state of the hardware safety systems, the state of the HV and LV, of the LHCb movable device (VELO) and the state of the timing distribution.

## III. CENTRALIZED READOUT SUPERVISION

As a design choice, the LHCb readout system is centrally controlled by a single Readout Supervisor [4]. The choice was made such that the Readout Supervisor has connection to all sub-systems of the readout architecture in order to:

- receive and distribute the global timing synchronous with the LHC timing
- receive and distribute the first-level hardware trigger decision
- generate and distribute synchronous and asynchronous commands for readout control
- interfacing to the LHCb ECS for real-time run management and activity configurations
- absolute timestamp and generate a description of each and every accepted event
- handle backpressure from the readout system and the software farm
- distribute events over the network in a dynamic way to reduce backpressure, network blockages and balance the processing of events at the software farm.

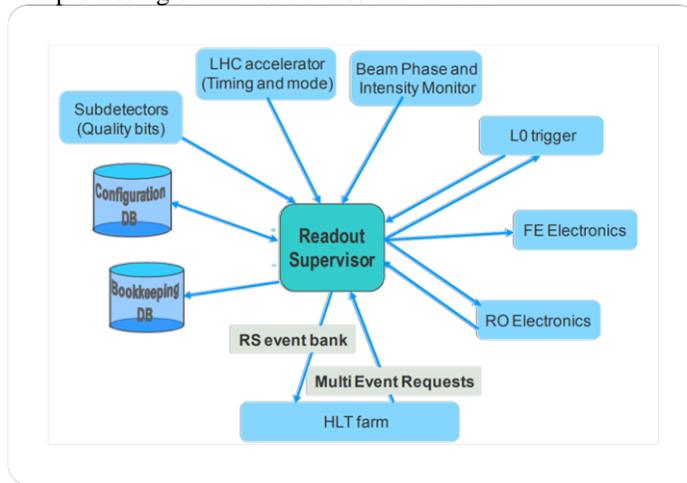

Fig. 5: The LHCb readout system is centrally controlled by a single Readout Supervisor with interfaces to all sub-systems of the readout architecture.

The choice of having a single centralized Readout Supervision allows for the highest level in flexibility in performing procedures and controlling the readout system:

- calibration and event type selection can be performed dynamically and based on easily changeable configuration
- global resets and synchronization mechanisms across the entire readout system allows to maintain synchronicity and ultimately a more smooth data taking from the operation point of view
- direct interface to the LHCb control system and to the framework for data exchange with the LHC accelerator allows to dynamically configure the Readout Supervisor based on data taking different conditions

As a practical example, the single Readout Supervisor can be programmed such that it contains the LHC filling scheme (this is the scheme that defines the position of the proton bunches in the LHC ring wrt to the associated bunch crossing identifier). This is loaded in a memory in the central processing FPGA. The information of the type of crossing - whether it is a colliding crossing or not - is used to tag and match a trigger decision to ultimately select the interesting events as well as create a report about the accepted event. In fact, such information can override the first-level trigger decision should this decision arrive at a bunch crossing that it is not considered interesting or should the Readout Supervisor be configured in a way to only accept a very specific type of trigger. The choice of implementing this centrally and in an FPGA allows for programmability of such functionalities and the highest level of flexibility of readout modes and recipes while keeping the data acquisition efficient.

## IV. POWERFUL TRIGGER SYSTEM AND REAL-TIME CALIBRATION AND ALIGNMENT

An even more effective selection and a higher signal purity of the relevant decay channels to the LHCb physics program can be achieved by a dedicated alignment and calibration procedure [5]. While this is quite commonly done in offline reprocessing, the novelty introduced by the LHCb experiment was to perform it in real-time – at the same time as online data taking. The selected events after the first stages of the software and hardware triggers are buffered on local disks and an automatic calibration and alignment of the detector is performed. This procedure enables the best possible calibration to be applied already at the software trigger level and provide a better performance in the trigger. A by-product of this is that the freshly collected and calibrated events are ready for analysis only few hours after they have been recorded by the detector. The infrastructure behind such operation is a now well-oiled machine and it involves a complex interplay between the control system, the timing system, the data acquisition system and the trigger system, and the actual calibration tasks.

The events used as "control" channels to generate a set of calibration/alignment constants are selected within the trigger system [6]. First the first-level hardware trigger selects ~1 MHz of events out of the pool of available ~40 MHz events at the LHC, then the central Readout Supervisor tags them and defines an event description with the most important information regarding the event that was selected. The Readout Supervisor also defines a common destination to each fragment of the event so that a software event builder can pack all the fragments in a single event in the same location. The newly built event is then passed to the high-level trigger software which applies a selection on the events and reduces the rate of events to ~125 kHz, by parking the data to local disks. Among these data, the control system monitors the collection of sub samples to feed the calibration tasks. When the minimal target amount of events is reached, the actual output of the online calibration and alignment of the detector is performed and finally the control system applies such changes to the trigger. This is schematically and simplistically shown in Fig. 6.

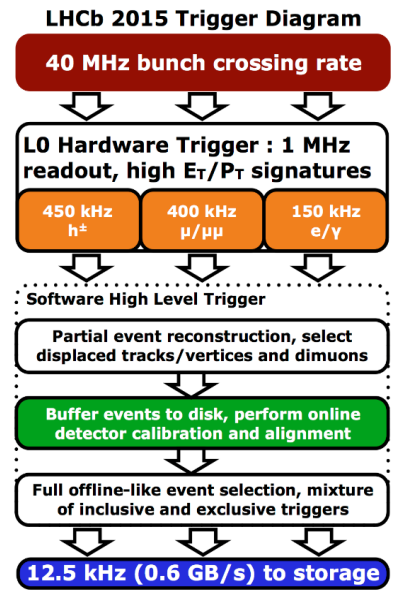

Fig. 6: Event flow and trigger diagram as it was implemented in the LHCb readout system.

While the whole procedure just described is completely automatized, the output of the online calibration is published to the shifters, who are in charge of checking that the automatic procedure was successful. An alarm signals to the shifters that the output is ready and that they should check the quality of such output, limiting the human interaction to the bare minimum.

This procedure must be redone every time running conditions change: for example, during data collection in *"PHYSICS"* conditions if the LHCb movable device (Vertex Locator, VELO) moves out of its data taking position to go into parking position, the VELO alignment task must be relaunched and its output must be taken into account in the trigger. Alternatively, if the LHC accelerator changes its mode of operation, moving out of the stable conditions of delivering physics. In this case, protection mechanisms are automatically initiated to put the voltages of the sub-detector in a safe state, stop the data taking and put the experiments in an idle state. This is done by the LHCb control system as described in the previous chapter and underlines even more how the choices described in the previous chapter play a fundamental role in efficient data taking.

## V. ONLINE LUMINOSITY, BEAM AND BACKGROUND MONITORING

The last major real-time aspect in operating the LHCb experiment is the online monitoring of beam, background and luminosity conditions. In fact, to fulfil the LHCb physics program, it is desired to obtain a homogenous dataset. This means that the events should all resemble each other in terms of number of tracks, number of primary and secondary vertices, multiplicities and occupancies. This is achieved by actively choosing the operating point in which the LHCb detector records data: this is normally obtained by monitoring the value of pileup at the LHCb interaction point – pileup is here defined as the average number of visible proton-proton interactions per bunch crossing and commonly referred to as µ. In LHCb the desired value of µ is 1.1 and such value of pileup is desired no matter what the conditions of the LHC accelerator are (i.e. how many colliding bunches, their intensity or their location around the ring, etc). Fig. 7 shows the average value of µ throughout the year 2017 and it shows how the LHCb experiment records data at a constant value of pileup over the course a full year.

To be able to maintain such stable conditions, the LHC beams are separate in the vertical plane while kept optimized in the horizontal plane in the LHCb interaction region. Such procedure of controlling the value of pileup while separating the LHC beams is called *luminosity levelling* and it is achieved by an interplay of all the real-time aspects previously described:

1. the LHCb ECS receives the information regarding the beams, such as the LHC filling scheme, the emittance and the state of the LHC accelerator

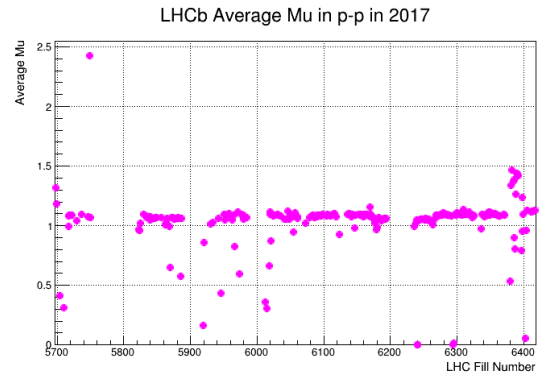

Fig. 7: Trend of the average LHCb µ value throughout the year 2017.

2. the LHCb Readout Supervisor is interfaced to the first-level hardware trigger and it receives the Minimum Bias triggers to count luminosity
3. a calibration procedure is regularly performed to obtain the value of processes cross-section to normalize the trigger rates to the corresponding value of luminosity
4. a software application (Fig. 8) obtains all of these information and computes the value of pileup
5. when the value of pileup differs from the target one, the application transmits a *levelling request* to the LHC control room via the exchange software framework [7]
6. the LHC separates the beams accordingly and a real-time feedback value of luminosity is provided at a rate of 1 Hz.

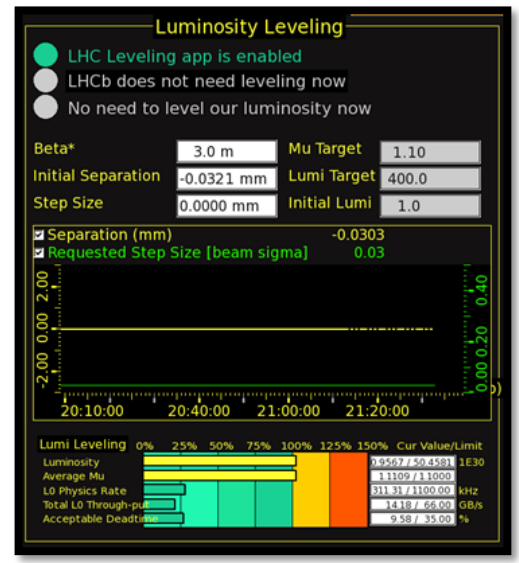

Fig. 8: The interface panel of the LHCb luminosity levelling application.

Due to the limited space here, only the luminosity levelling specific example is given in the topic of online beam and luminosity monitoring, nevertheless other aspects fall in this category: experiment protection and safety with respect to bad background conditions, beam timing and its relation to the clock distribution in the LHCb experiment and correlation of vacuum conditions with respect to the beam induced background conditions.

## VI. An LHCb Upgrade

Although the current LHCb experiment is performing extremely well and with high efficiency [8], the first-level hardware trigger is in fact the limiting factor in obtaining the best yield with respect to the amount of data collected.

The value of $\mu = 1.1$ was in fact chosen such that the first-level hardware trigger is at its limits before saturating for the hadronic channels. The LHCb experiment could in principle work at a higher value of pileup (hence higher luminosity which in turn means collecting more data), but the needs of reducing the data bandwidth from the full 40 MHz bunch crossing rate at the LHC to the 1 MHz first-level trigger rate means that at higher values of pileup, the first-level hardware trigger would need to cut much harder in its channel, effectively reducing the output yield in some physics decays.

The solution of this problem is to entirely remove the first-level hardware trigger and record data in a completely trigger-less fashion. Recorded data are fully transmitted to a flexible trigger system where filtering is then done by having the full event at its disposal, while keeping all of the real-time aspects that are described in this paper. The upgrade of the LHCb detector comprises the replacement of the entire Front-End, Back-End and DAQ electronics and up to 90% of detector channels and it is foreseen to happen in the period 2019-2020 to be ready for data taking starting in 2021 [9].

## VII. Conclusion

In this paper, the major real-time aspects in the LHCb experiment have been reviewed. These aspects are well-oiled, performant and novel solutions that have been developed over the past decade of operation of the LHCb experiment at the CERN LHC.

Such aspects comprises automatized operations, a centralized Readout Supervision, a powerful and flexible trigger system, online real-time calibration and alignment procedures, beam, background and online luminosity monitoring and the way all of these aspects interact with each other. Such implementations allowed to produce efficient and reliable operations and data taking at the LHCb experiment while still being operated by a pool of ~300 non-expert shifters and in a consistent way with the LHC accelerator.

Lastly, the potential of the LHCb experiment by combining its successful approach in real-time aspects and the foreseen upgrade detector with no hardware trigger are going to be a major tool in the quest for New Physics at the LHC.